\newif\ifsuppinfo
\suppinfotrue  

\documentclass[pdflatex,sn-nature, iicol]{sn-jnl}

\usepackage{graphicx}%
\usepackage{multirow}%
\usepackage{amsmath,amssymb,amsfonts}%
\usepackage{amsthm}%
\usepackage{mathrsfs}%
\usepackage[title]{appendix}%
\usepackage{xcolor}%
\usepackage{textcomp}%
\usepackage{manyfoot}%
\usepackage{booktabs}%
\usepackage{algorithm}%
\usepackage{algorithmicx}%
\usepackage{algpseudocode}%
\usepackage{listings}%

\usepackage[normalem]{ulem}

\theoremstyle{thmstyleone}%
\theoremstyle{thmstyletwo}%

\theoremstyle{thmstylethree}%

\DeclareMathOperator*{\argmax}{arg \, max}
\DeclareMathOperator*{\minimum}{min}

 \newcommand{\vone}[1]{{}}
 \newcommand{\vtwo}[1]{{#1}}

\usepackage{bibunits}

\unnumbered

\begin{document}

\title{Interpolating many-body wave functions for accelerated molecular dynamics \vtwo{on the near-exact electronic surface}}

\author*[1,2]{\fnm{Yannic} \sur{Rath}}\email{yannic.rath@npl.co.uk}
\affil[1]{\orgname{National Physical Laboratory}, \orgaddress{\city{Teddington}, \postcode{TW11 0LW}, \country{United Kingdom}}}

\author*[2]{\fnm{George H.} \sur{Booth}}\email{george.booth@kcl.ac.uk}

\affil[2]{\orgdiv{Department of Physics and Thomas Young Centre}, \orgname{King's College London}, \orgaddress{\street{Strand}, \city{London}, \postcode{WC2R 2LS}, \country{United Kingdom}}}



\abstract{
    While there have been many developments in computational probes of both strongly-correlated molecular systems and machine-learning accelerated molecular dynamics, there remains a significant gap in capabilities in simulating accurate non-local electronic structure over timescales on which atoms move. We \vtwo{develop an approach to bridge these fields with a practical interpolation scheme for} the correlated many-electron state through \vtwo{the space of atomic configurations}, whilst avoiding the exponential
    complexity of these \vtwo{underlying electronic} states. With a small number of accurate correlated wave functions as a training set, we demonstrate provable convergence to near-exact potential energy surfaces
    for subsequent dynamics with propagation of a valid many-body wave function and inference of its
    variational energy whilst retaining a mean-field computational scaling. This represents
    a profoundly different paradigm to the direct interpolation of \vtwo{potential energy surfaces} in
    established machine-learning approaches. We combine this with modern electronic structure approaches to systematically resolve molecular dynamics \vtwo{trajectories and converge thermodynamic quantities with a high-throughput of several million interpolated wave functions with explicit validation of their accuracy from only a few numerically exact quantum chemical calculations. We also highlight the comparison to} traditional machine-learned potentials or dynamics on mean-field surfaces.
}

\maketitle

\begin{bibunit}[sn-nature]


    The quantum fluctuations of interacting electrons represent the critical interaction between atoms which underpin all atomic bonding, dynamics and reactivity. Computational approaches for systems with strongly interacting electrons have undergone a number of major developments in recent decades, as emerging methods enable a description of correlated electronic structure for ever larger and more realistic systems with unprecedented accuracy~\cite{PhysRevX.7.031059, PhysRevX.10.011041, doi:10.1021/acs.jpclett.0c02621}. These modern approaches across both chemical and materials science include those based on tensor networks~\cite{10.1063/1.4905329,10.1063/5.0050902, Larsson2022TheCD, doi:10.1126/science.abm2295, Li2019}, stochastic methods~\cite{Booth2012TowardsAE, 10.1063/5.0150706, Ren2023, doi:10.1021/acs.jctc.1c01162}, selected configuration interaction~\cite{KOCH1993193, 10.1063/1.5063376, 10.1063/1.1679199} and machine-learning inspired wave function ansatze~\cite{Choo_2020, https://doi.org/10.48550/arxiv.2109.12606, zhaoScalableNeuralQuantum2022, https://doi.org/10.48550/arxiv.2301.03755, Yang2020, PhysRevB.107.205119, bortone2023impact, PhysRevResearch.2.033429, Hermann2019DeepneuralnetworkSO, gerard2022goldstandard}. \vtwo{This has} allowed for the near-exact solution to the quantum many-\vtwo{electron} problem in these systems, providing high-accuracy insights for a few fixed atomic configurations, but have in general had little or no impact on our understanding of the physics and chemistry on the timescales of atomic and molecular motion.

    The reasons for this are obvious; while a small number of single point calculations with fixed \vtwo{nuclei} are possible, the different timescales of atomic dynamics and electronic quantum fluctuations mean that on the order of at least thousands of sequential electronic structure calculations are required. This is essential to propagate the atoms in molecular systems to relevant timescales, entailing generally prohibitive computation expense for these high-accuracy methods. This is particularly challenging for these emerging methods which can lack a `black-box' use, requiring care to ensure reliable convergence at each point, while often also lacking analytic atomic forces to propagate the nuclear coordinates in time~\cite{doi:10.1021/acs.chemrev.9b00496}. \vtwo{Important developments have been made in recent years in extending the application of established ground-state quantum chemical models to atomic dynamics~\cite{doi:10.1021/jz401931f,10.3389/fmats.2015.00029,10.1063/1.4941091,C4CP05192K,Hutter2018,10.1063/5.0212274}, while `active space' methods are also increasingly widely used for stronger correlation or excited state molecular dynamics~\cite{doi:10.1021/acs.chemrev.7b00577}.} However, the additional cost of these approaches have meant that `{\em ab-initio} Born-Oppenheimer molecular dynamics' (AI-BOMD), where the atoms are classically propagated according to the potential energy surface of the electrons, is almost synonymous with a more empirical density functional description of the electronic structure
    which lacks systematic improvability and has many well documented deficiencies~\cite{PhysRevLett.55.2471,doi:10.1073/pnas.0500193102,marx2000ab}. These include an over-stabilisation of delocalized electronic states, as well as often inaccurate descriptions of dispersion forces, transition states or bond breaking among others~\cite{Cohen2012ChallengesFD}. These are critical parts of the phase space in real chemical dynamics, and the acute need for more reliable potential energy surfaces which build on the developments in accurate electronic structure is clear.

    The most widespread and successful resolution to this need has come from the `machine learning' of force fields~\cite{Kabylda2023EfficientID, Fu2022ForcesAN, Unke2020MachineLF,Gkeka2020MachineLF}. These interpolate across chemical space between accurate single-point estimates of the electronic energy, based on local descriptors of the environment of each atom~\cite{doi:10.1021/acs.chemrev.1c00021}. While this approach to straddling the electronic and atomic timescales has been arguably one of the most successful contributions of machine learning to quantum-level simulations to date, it is not without its own drawbacks~\cite{Kulik_2022}. In particular, the local nature of the descriptors can lead to difficulty describing long-range interactions \cite{doi:10.1021/acs.jctc.3c00704}, as well as `holes' where non-variational inferred energy estimates can lead to a collapse in the statistical sampling of phase space to these unphysical minima~\cite{doi:10.1021/acs.jctc.1c00647}. On a more fundamental level, since these approaches integrate out the electronic structure, there is in general no fundamental electronic variable at each sampled point (such as the wave function), meaning that the electronic properties which can be extracted are limited to the ones which correspond to the model definition. If the evolution of e.g. the dipole moment or charge distribution across a trajectory was desired this would not be accessible from a force field, and extensions to non-adiabatic effects are also far from straightforward in this framework, noting however significant recent research in these directions \cite{Schtt2019UnifyingML, Fedik2022ExtendingML,doi:10.1021/acs.jpclett.0c00527,D3FD00113J}.

    We take a different perspective and show that rather than interpolating observables such as the potential energy, we can instead interpolate the many-body electronic wave function itself through \vone{chemical space}\vtwo{the phase space of molecular conformations}. Importantly, despite the many-body wave function of each training point being in general exponentially complex, inference of properties from the model can be achieved in a scaling which is the same as (hybrid) density functional theory, rendering this a practical scheme.
    This decouples the unfavourable scaling of high accuracy single point electronic structure calculations from the evaluation of the interpolated potential energy surface, and thus allows for the use of these electronic structure methods for molecular dynamics on realistic timescales.
    We show that the resulting potential energy surfaces and molecular dynamics are systematically improvable to near-exactness via interpolation between highly accurate training configurations. Since a valid correlated many-electron state is propagated through \vone{chemical space} \vtwo{the sampled phase space}, this paradigm enables all electronic properties of interest to be simultaneously accessible within the same model, without relying on local or low-rank descriptors. Furthermore, since the energy is computed as a rigorous quantum expectation value over this inferred state, it provides a fully variational potential energy estimate (precluding `holes') for all atomic configurations, allows for clear evidence of systematic improvability to exactness as the training set is enlarged, an inductive bias of the model away from poorly described regions of phase space, and simple access to analytic atomic forces of the model for efficient propagation of dynamics.

    We combine this approach for interpolating wave functions with modern density matrix renormalization group (DMRG) methods, allowing convergence of the strongly correlated potential energy surfaces to near exactness \vtwo{within the employed basis}~\cite{10.1063/1.4905329,10.1063/5.0050902}. We demonstrate that this can provide a fully correlated electronic description of reactive molecular dynamics beyond traditional parameterized or machine-learned force fields, \vtwo{and ensembles of thermalized trajectories for equilibrated quantities} over time scales which would be inaccessible without this acceleration. We show this can result in qualitative differences in behavior for a number of proto-typical molecular dynamics simulations compared to both density functional and traditional machine-learned force field approaches~\cite{10.1063/5.0160898}. Finally, we \vtwo{compute both thermalized expectation values from canonical emsembles and reactive high-energy dynamical trajectories} on a near-exact potential surface for the Zundel cation, a key intermediate for the Grotthuss mechanism for hydrogen diffusion through aqueous solutions~\cite{Agmon1995TheGM,Cukierman2006EtTG,10.1063/1.4941091}. \vtwo{With explicit validation of the accuracy of the surface, we compare the dynamics to both density functional theory results and other quantum chemical methods for both for structural and electronic quantities, highlighting marked differences which can result from the quality of the surface.}

    \section*{Results and discussion}
    \subsection*{Interpolating wave functions}\label{sec:interpolation}

    We first consider how to interpolate a single many-electron wave function between two different atomic configurations. We assume that we have an exact (FCI) correlated many-electron state defined within an atom-centered basis set of $L$ functions, for a specific set of atomic coordinates ${\bf R}$~\cite{KNOWLES198975}. This wave function is a linear superposition over exponentially many electron configurations (Slater determinants) spanning the Hilbert space, as
    \begin{equation}
        |\Psi({\bf R}) \rangle = \sum_{n_1, n_2, \dots, n_L} C_{n_1, n_2, \dots , n_L} |n_1, n_2, \dots , n_L \rangle , \label{eq:manybodywfn}
    \end{equation}
    where $C_{n_1, n_2, \dots, n_L}$ is the rank-$L$ tensor of probability amplitudes over the electronic configurations, and $n_i$ indexes the four local Fock states of the $i^\textrm{th}$ orbital; either unoccupied, spin-up, spin-down, or doubly occupied with electrons for each orbital. In general, both the probability amplitudes, and the single-particle orbitals defining each electronic configuration $|n_1, n_2, \dots , n_L \rangle$ will change with atomic configuration ${\bf R}$. However, we aim to represent an approximation to the correlated electronic state at a {\em different} \vone{chemical}\vtwo{atomic} configuration (and therefore electronic Hilbert space) with the {\em same} tensor of probability amplitudes over these electronic states. We exploit the fact that the properties of the exact state are invariant to orthogonal rotations of the single-particle orbitals, but that the probability amplitudes themselves will vary with this choice. Therefore, to enable transferrability between chemical environments, we seek a choice of orbital representation in which the probability amplitudes of the exact many-electron state change {\em least} between atomic configurations of interest.

    A plausible choice is a basis of local atomic-like functions, appealing to the fact that a large portion of the electronic fluctuations among atomic-local orbitals will remain qualitatively similar as atoms are moved by small amounts. Similarly, regions of similar chemical bonding will also have common features in their probability amplitudes defining e.g. covalent fluctuations between neighboring atoms~\cite{mejutozaera2023quantum}. However, for reasons which will become clear, we also require that the orbitals represent an orthonormal set for all atomic configurations. To ensure this, while (in a least-squares sense) optimally preserving this atomic-like character of the orbitals, we symmetrically (Löwdin) orthonormalize the atomic-orbital basis (see Methods), defining orthonormal `SAO' orbital sets for each atomic configuration~\cite{10.1063/1.1747632, https://doi.org/10.1002/qua.981}. We can then choose to interpolate the state (and all resulting properties) between atomic configurations without re-optimizing the many-electron state by simply transferring the probability amplitudes, while ensuring the consistent SAO basis definition.

    This simple approach is limited, since the many-body amplitudes will in general change as the atoms move. However, we can generalize the state while retaining a valid wave function by linearly combining probability amplitude tensors in this transferable SAO representation from a larger `training' set optimized at {\em other} atomic configurations.
    We then variationally optimize the relative contributions of each of the $N$ training states for any test atomic configuration. This is achieved in closed form as the diagonalization of a generalized eigenvalue problem in the basis of the training states. By projecting the Hamiltonian at the desired test geometry $\mathbf{R}$ into this many-body basis, we get
    \begin{equation}
        \boldsymbol{\mathcal{H}}(\mathbf{R}) \mathbf{X}(\mathbf{R})=\mathbf{E}(\mathbf{R}) \, \boldsymbol{\mathcal{S}} \mathbf{X}(\mathbf{R}) , \label{eq:geneig}
    \end{equation}
    with the eigenvectors, $\mathbf{X}(\mathbf{R})$, giving the amplitudes of the training states defining the interpolated wave functions at the test geometry, with inferred energy spectrum $\mathbf{E}(\mathbf{R})$. The electronic Hamiltonian of the test geometry, $\boldsymbol{\mathcal{H}}(\mathbf{R})$, is \vtwo{found by projecting the Hamiltonian operator into the many-body} basis defined by the fixed probability amplitudes of the \vone{many-body} training states. \vtwo{This can be found in compact form as}
    \begin{align}
        \mathcal{H}_{ab} & = \sum_{ijkl} \sum_{\mathbf{n} \mathbf{n}'} C_{\mathbf{n}}^{(a)*} C_{\mathbf{n}'}^{(b)} \langle \mathbf{n} | {\hat c}^\dagger_i {\hat c}^\dagger_j {\hat c}_l {\hat c}_k | \mathbf{n}' \rangle \, K_{ijkl}(\mathbf{R}) \nonumber \\
                         & = \sum_{ijkl} \Gamma^{ijkl}_{ab} \, K_{ijkl}(\mathbf{R}), \label{eq:subspaceh}
    \end{align}
    where $\mathbf{n}$ denotes the many-electron configurations in the SAO basis of the test geometry, with $|\mathbf{n}\rangle \equiv |n_1, n_2,\dots,n_L\rangle$, and with $C_{\mathbf{n}}^{(a)}$ and $C_{\mathbf{n}'}^{(b)}$ the fixed SAO probability amplitudes of the training \vone{points}\vtwo{wave functions} at atomic geometries $a$ and $b$ respectively. This definition therefore implicitly transfers the probability amplitudes between the Hilbert spaces of the training and test geometries. The $K_{ijkl}(\mathbf{R})$ tensor is the two-electron reduced Hamiltonian defined in the SAO basis of the test geometry (given explicitly in the Methods section) with second-quantized Fermionic operators acting in this basis as ${\hat c}^{(\dagger)}$ shown.

    Since the Hamiltonian is a sum of only two-electron interactions, the contraction over the exponential many-electron configurations, $\mathbf{n}$, is performed for all pairs of training states to give the transition two-body density matrices $\boldsymbol{\Gamma}^{ijkl}$ defined in Eq.~\ref{eq:subspaceh}. Crucially, since the training probability amplitudes are defined not to change with test geometry, this contraction is only performed once on the training states, {\em not} for each test geometry. The construction of the subspace Hamiltonian at a test geometry therefore only requires the $\mathcal{O}[L^4]$ contraction of Eq.~\ref{eq:subspaceh}, with only \vtwo{the} $K_{ijkl}(\mathbf{R})$ \vtwo{term} changing with test configuration. The overlap between the training states, $\boldsymbol{\mathcal{S}}$ of Eq.~\ref{eq:geneig}, can similarly be precomputed during the training phase, as
    \begin{equation}
        \mathcal{S}_{ab}=\sum_{\mathbf{n}} C_\mathbf{n}^{(a)*} C_{\mathbf{n}}^{(b)} ,
    \end{equation}
    due to the orthonormality of the SAO basis at all test geometries. This ensures that the overlaps of the training states do not change with geometry, despite the physical training wave functions changing with atomic rearrangements as they transform between Hilbert spaces. In this way, the exponential complexity of the many-electron states are completely avoided in the inference of wave functions at new test points, by representing the training states in the polynomially-compact tensors $\boldsymbol{\Gamma}^{ijkl}$ and their overlaps. \vtwo{The inference of the model requires a computational scaling of non-iterative $\mathcal{O}[N^2L^4]$ after the density matrices of the training states have been precomputed in the training stage -- the same formal scaling with system size as traditional (hybrid) density functional theory. This scaling for evaluation of the model at test geometries could also be further lowered with factorizations exploiting the low-rank nature of the $\Gamma_{ab}^{ijkl}$ tensors~\cite{doi:10.1021/acs.jctc.7b00605, 10.1063/1.5047207, doi:10.1021/acs.jctc.8b00780,doi:10.1021/acs.jctc.3c00851}.}

    The lowest energy eigenvector from the diagonalization of this Hamiltonian (whose dimensionality scales only with the number of training points and is independent of system size) defines the specific variationally optimal linear combination of training probability amplitudes for the state, which can subsequently be used to predict any electronic property at this test geometry.
    Due to the variationality we have the desirable properties that the inferred state at a point which coincides with a training geometry must necessarily be exact, as well as the fact that each additional training point must necessarily lower the inferred energy towards the exact electronic solution across {\em all} possible test atomic configurations, assuming linear independence. In this way, the method more closely resembles a reduced order method than a machine learning model, \vtwo{where we define the Hamiltonian in a subspace defined by a fixed set of many-body vectors taken from training wave functions at different geometries}, yet both are useful viewpoints. Due to the fixed training amplitudes across geometries, as well as the variational optimization of the model, computing analytic atomic forces from the inferred state is also straightforward \vone{via the Hellmann-Feynman theorem} (see Methods), ensuring a particular relevance of this acceleration in molecular dynamics applications.

    \begin{figure*}
        \includegraphics{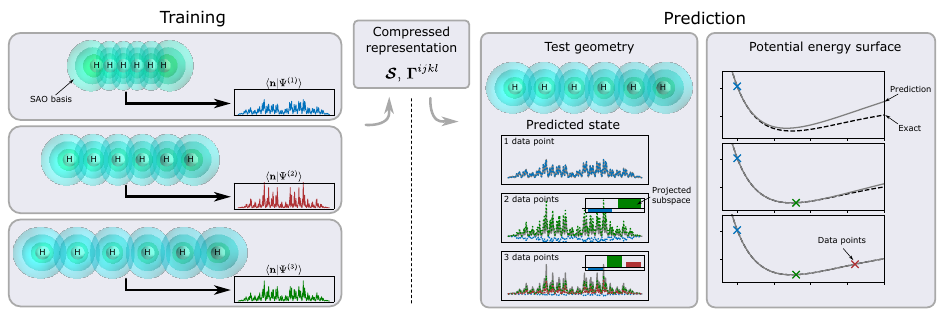}
        \caption{Schematic overview of the proposed `Eigenvector Continuation' scheme for the high accuracy prediction across \vone{chemical}\vtwo{conformational} space from few \emph{ab initio} data points. Utilizing training states in a Löwdin atomic orbital (SAO) basis, states are predicted for new molecular geometries by diagonalization in the subspace of the training states, allowing for a fast variational prediction of highly accurate potential energy surfaces and force fields via the compact one- and two-body transition density matrices between states. The scheme is exemplified for symmetrically stretched chains of $6$ hydrogen atoms in a minimal basis set (STO-6G) using one, two, and three training points. $\boldsymbol{\mathcal{S}}$ and $\boldsymbol{\Gamma}^{ijkl}$ represent the geometry-agnostic overlap and two-body transition density matrices between training states. $\langle \mathbf{n} | \Psi^{(a)} \rangle$ shows the value of the (exponentially many) wavefunction amplitudes in the SAO basis for the $a^\mathrm{th}$ training state, and the test geometry shows the combination of training states used to define its own SAO probability amplitudes.}
        \label{fig:schematic}
    \end{figure*}

    This approach also builds on the perspective of `eigenvector continuation' which was recently introduced in both nuclear physics and condensed matter lattice models, where an eigenstate is analytically continued to different parts of the phase diagram~\cite{PhysRevLett.121.032501,PhysRevC.101.041302,duguet2023eigenvector,KONIG2020135814,PhysRevC.107.064316,DRISCHLER2021136777,10.1063/5.0141145,D4FD00062E}. Even more recently this was extended to simple {\em ab initio} quantum chemistry applications, with a related scheme to the one proposed~\cite{mejutozaera2023quantum}. However, a crucial difference was the use of a non-orthogonal atomic basis, which necessitated evaluation of the test point Hamiltonian directly from the many-body states. This retained the exponential complexity of the many-body state for inference at each test geometry, which is avoided here. The method of Mejuto-Zaera {\em et. al.} was therefore presented instead as an approach for {\em quantum} computers, where many-body unitary operations can be applied in polynomial complexity. In contrast, the SAO basis for the interpolation formally breaks this requirement, ensuring the approach is amenable to classical computation in the predictions at test points with tractable mean-field computational cost.

    A simple example of the scheme is shown in Fig.~\ref{fig:schematic}, for the symmetric stretch of a chain of six hydrogen atoms, with up to three atomic displacements considered in the training set. The compressed intermediate representation of the overlaps and transition density matrices between the training states in the SAO basis are shown, enabling variationally optimal predictions across the whole potential energy surface as linear combinations of the many-body training basis transformed between geometries. The predictions are found to converge to near exactness for this system with only three training points, with a guarantee of smoothness on adiabatic surfaces, and exactness at any training point geometry.

    This approach is invariant to translation and rigid body rotations of the chemical system, provided a consistent ordering of the SAO representation is maintained, which is straightforward to achieve. However, this is not trivial for atomic permutations or point group symmetries which would change the SAO ordering, and hence probability amplitudes of the state definition. Furthermore, in contrast with building a force field based on local descriptors, the inference requires the same dimensionality Hilbert space for the electronic state, necessitating that the training and prediction points are taken from the same sized system. \vtwo{This is a significant difference to force field approaches with local representations which allow for scaling the system size after training, ensuring a different scope of applicability to the proposed approach~\cite{doi:10.1021/acs.chemrev.1c00021}.} Future work will look to relax this constraint.


    \begin{figure}
        \includegraphics{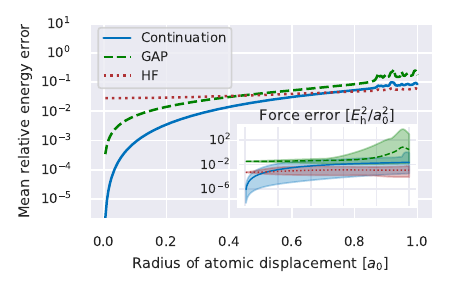}
        \caption{Mean relative energy error of the prediction for distorted ten-atom hydrogen chains against the absolute displacement of each atom from the equilibrium position. For each realization, a distorted chain was created by moving each atom from their position in the equilibrium geometry by a fixed displacement with a random direction. The comparison includes predictions from Hartree-Fock (red, dotted), the Gaussian approximation potential framework (green, dashed), as well as the variational continuation scheme from $5$ training states of the symmetrically stretched chain (blue, solid). Each data point corresponds to the mean over 1000 randomly generated geometries. The inset shows the mean squared force error obtained with the three methods, where the shaded area denotes the range of the errors over the random realizations. The training set of equidistant one-dimensional geometries include the equilibrium length, with an interatomic distance of $\approx 1.79 \, a_0$, as well as the $4$ symmetric stretches of the atoms where the inter-atomic distance was increased and decreased by $0.5 \, a_0$ and $1 \, a_0$.}
        \label{fig:PES_H10}
    \end{figure}

    While the proof-of-principle in Fig.~\ref{fig:schematic} demonstrates excellent accuracy with few training points, it is also interpolation within a simple one-dimensional phase space of geometries. We now compare to a far larger phase space, composed of averaging the errors in both the inferred energy and analytic forces on the atoms over randomly oriented three-dimensional displacements of each atom from a ten-atom linear hydrogen chain. This provides an exponentially large phase space of distorted chain configurations to test, where the radius of the displacements of each atom can be used to control the magnitude of the geometric distortions from the parent linear chain \vtwo{from which the training data is obtained}. Only five training points from the symmetric stretch of the equidistant linear chain are used. We consider the increase in error as the magnitude of the displacements are increased in Fig.~\ref{fig:PES_H10}, as the test configurations move further from these training samples.
    We also compare these errors to a Gaussian approximated potential (GAP); a widely used machine-learning approach based on Gaussian process regression in a space of local descriptors from the superposition of atomic potentials~\cite{PhysRevLett.104.136403,10.1063/5.0160898}. This models a force field directly from the same training energies, but results in a materially larger error for the energy and forces over all displacements. We note that five points would generally be a very small training set for GAP, and that improved techniques to directly train on the forces of the training data themselves or improved model definitions were not used~\cite{PhysRevB.95.214302,Kozinsky21}. 
    
    Nevertheless, a demonstration that inferring the wave function amplitudes themselves can outperform traditional machine-learning inference of the properties directly is noteworthy. Furthermore, we compare to Hartree--Fock theory (HF), which neglects all correlated electron effects and has the same computational scaling as the inference of the proposed `eigenvector continuation' scheme. This is also significantly worse at small distortions of the chain, though outperforms the largest distortions which are far from the training geometries and deep in the extrapolation regime.

    \subsection*{Bridging timescales}\label{sec:results}

    While it is easy to envisage many applications of an interpolation scheme for accurate correlated electronic structure, an obvious target is Born-Oppenheimer molecular dynamics (MD)~\cite{marx2000ab}. In particular, the variationality of the scheme allows for systematic and quantifiable improvability to the exact solution of the electronic Schr{\"o}dinger equation in the inferred potential energy surface at each geometry, while retaining a mean-field scaling with respect to the timescales which can be accessed. To access larger systems and basis sizes we also turn to modern electronic structure approximations for the evaluation of training states. In particular, we use the density matrix renormalization group (DMRG) to obtain training states with controllable accuracy to exactness~\cite{10.1063/1.4905329,10.1063/5.0050902}. These DMRG calculations can either be performed directly in the SAO basis or the state rotated into this basis after optimization, in advance of computation of the required overlaps and transition density matrices between the training states. As an alternative to DMRG, we are also able to approximate the training states by restricting the space of correlations to a low-energy complete active subspace (CAS) selected from the low-energy orbitals of a mean-field calculation~\cite{10.1063/5.0042147}.

    \begin{figure}
        \includegraphics{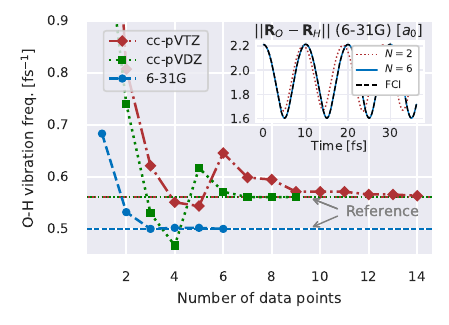}
        \caption{Predicted frequency of the $a_1$ symmetric stretch of a water molecule obtained from MD trajectories starting from a stretched initial configuration, predicted with different numbers of training data geometries for the continuation. We simulate the system in increasingly large 6-31G (blue, dashed), cc-pVDZ (green, dotted) and cc-pVTZ (red, dash-dotted) basis sets where the larger two bases use training data restricted to a CAS space of $4$ electrons in $8$ Hartree--Fock orbitals. Horizontal lines give reference values from MD results on a FCI surface in the 6-31G basis, and CAS simulations in the cc-pVDZ and cc-pVTZ basis. The inset shows the oxygen-hydrogen distance over the MD trajectory in the 6-31G basis, as obtained from continuation with $N=6$ (blue, solid) and $N=2$ (red, dotted) training points, as well as the reference trajectory from FCI (black, dashed).}
        \label{fig:H2O_vibration}
    \end{figure}

    We consider these approaches for constructing training states and the subsequent MD of a water molecule in increasing basis sets in Fig.~\ref{fig:H2O_vibration}. In particular, we consider convergence of the predicted vibrational frequency of the $a_1$ symmetric stretching mode as the number of training points increases. For the smallest basis, we find the full vibrational dynamics converge with just three training points, where we can compare directly to exact FCI calculations of the dynamics. As we increase the basis, FCI is intractable and we restrict the training to a CAS of low-energy orbitals, where the number of training points required grows modestly to seven and thirteen training points in a cc-pVDZ and cc-pVTZ basis respectively. While we compare to CASCI for these larger bases for a reference estimate, we note that the inferred energies are necessarily variationally {\em below} these approximate training states. \vtwo{Furthermore, any discontinuities in the potential energy which arise commonly due to abrupt changes in the in the active space character are necessarily removed in the interpolated surface. Both of these features are elaborated on in} the extended data.

    It was found important to develop an active learning scheme for the selection of appropriate atomic configurations to include in the training data for rapid convergence. In Ref.~\cite{mejutozaera2023quantum}, the energy variance was motivated as an appropriate measure for the inclusion of data points, however this is impractical in the current lower-scaling scheme as it would require the evaluation of higher-body transition density matrices between training states. Instead, we consider the addition of training points which will maximize the improvement in the MD trajectories while respecting the invariances in the model predictions. This is performed by selecting the point on the trajectory where the Hamiltonian operator in the SAO basis, $K_{ijkl}(\mathbf{R})$, has changed most (in a least squares sense) compared to the Hamiltonians employed to generate the current training data set. Since the probability amplitudes are uniquely defined by this Hamiltonian, it is a suitable measure for the addition of new data points. Furthermore, due to the variationality of the method, it is guaranteed that the potential energy with the enlarged training data will be equal or lower to the previous predictions, across the whole trajectory. This can therefore be used as a rigorous metric for the systematic convergence of the potential energy surface for the MD, with more details in the Methods section. We consider the potential energy surface over the whole MD simulation fully converged when the maximum reduction in energy for any point over the whole trajectory is less than $1 \, mE_\mathrm{h}$ for two consecutive increases in the data set size.

    \begin{figure}
        \includegraphics{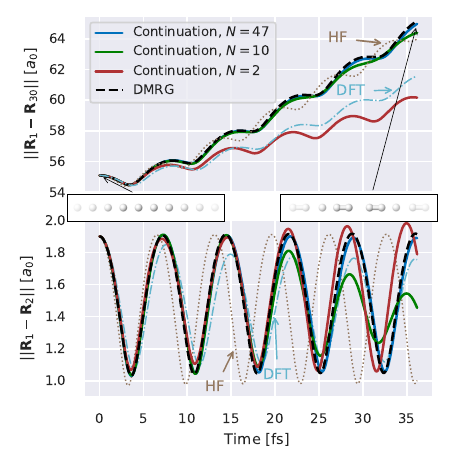}
        \caption{Molecular dynamics of a $30$ atom hydrogen chain in a STO-6G basis from an initially symmetrically stretched geometry. The panels report the distance between two of the hydrogen atoms (top panel: first and last atom in the chain, bottom panel: first and second atom) against the simulation time, showing that the first two hydrogen atoms form a stable vibrating dimer while the chain lengthens. The trajectories were obtained from the eigenvector continuation with $N=47$ (blue\vone{, solid}), $N=10$ (green\vone{, dashed}) and $N=2$ (red\vone{, dotted}) DMRG data points, together with the trajectories from DMRG (black, dashed), Hartree-Fock (\vone{grey, dashed}\vtwo{brown, dotted}) and density functional theory with a PBE exchange correlation (\vone{grey, dash-dotted}\vtwo{light blue, dash-dotted}) potential energy surfaces. Additional snapshots depict the initial and final hydrogen chain arrangements obtained from the converged eigenvector continuation ($N=47$).}
        \label{fig:H_chain_dimerization}
    \end{figure}

    A semi-infinite symmetric one-dimensional chain of hydrogen atoms has emerged as a paradigmatic benchmark system of strongly correlated electronic structure in recent years, as a platform towards larger {\em ab initio} and extended systems. Almost all modern electronic structure methods have been applied to the system with varying success, and it has motivated further developments in both theory and understanding of its unexpectedly rich phase diagram~\cite{PhysRevX.7.031059,PhysRevX.10.031058}. While the symmetric stretch of this system has been considered extensively via single-point electronic structure, its full dynamics at this level has not. In Fig.~\ref{fig:H_chain_dimerization}, we release the atoms to dynamically move on a tightly-converged ground state surface of the DMRG-trained continuation scheme, starting from a $\sim $10\% symmetric stretching of thirty atoms equally from the symmetric equilibrium structure. We find that along with the vibrations of the bonds, the atoms rapidly dimerize and separate, with the overall length of the chain increasing approximately linearly with time. We are able to converge the dynamics of this dimerization and dissociation (albeit in a minimal basis) \vtwo{to the equivalent explicit DMRG AI-BOMD} with only a small number of single-point training DMRG calculations. We note that in comparison, DFT-based AI-BOMD significantly underestimates the rate of dimerization of the chain, while Hartree--Fock theory conversely results in a bond for the hydrogen dimers which is too stiff, demonstrating the importance of an accurate treatment of the electronic correlations in the dynamics. 

    \subsection*{\vtwo{Towards chemical accuracy for realistic (thermo)chemistry}}

    We consider the feasibility of converging faithful \vtwo{thermodynamic quantities} and reactive chemistry on a near-exact potential energy surface for the gas-phase dynamics of a Zundel cation, comprising a water molecule and hydronium ion - a system whose intricate potential energy surface poses a challenging test case for novel numerical techniques, yet is particularly important for the understanding of proton diffusion in aqueous solution~\cite{D2SC03189B,Huang2005AbIP, McCoy2005FulldimensionalVC, KondatiNatarajan2015RepresentingTP, Heindel2018BenchmarkES, 10.1063/1.4941091}. 
    \vtwo{We first consider a statistical ensemble of $500$ different trajectories, starting from the same geometry (taken from Ref.~\cite{10.1063/1.1834500}), and sampling initial velocities from a Maxwell--Boltzmann distribution at $298.15 \, \mathrm{K}$. The BOMD was propagated under NVT conditions to thermalize according to a Berendsen integration scheme~\cite{berendsen1984molecular}. We consider $N=60, 80$ and $100$ single-point DMRG training configurations to observe the convergence of the thermodynamically equilibrated properties in a 6-31G basis. Each ensemble of trajectories at one of these training number involved $5 \times 10^6$ potential energy and force evaluations, which would be out of reach with a brute-force DMRG approach, but required a relatively modest 7,500 CPU hours for the propagation of the full ensemble. Nevertheless, we can explicitly verify convergence to the accuracy of the underlying DMRG by a validation of the `test error' via additional DMRG calculations for sampled geometries along the trajectories. The achieved test error, shown in the extended data, demonstrates that the PES is well below chemical accuracy of the exact potential energy surface within the employed basis as the thermal equilibrium is approached, reaching relative correlation energy errors below that of both CCSD and CCSD(T) -- the `gold standard' of quantum chemistry \cite{RevModPhys.79.291}.}
    

    \begin{figure}
        \includegraphics{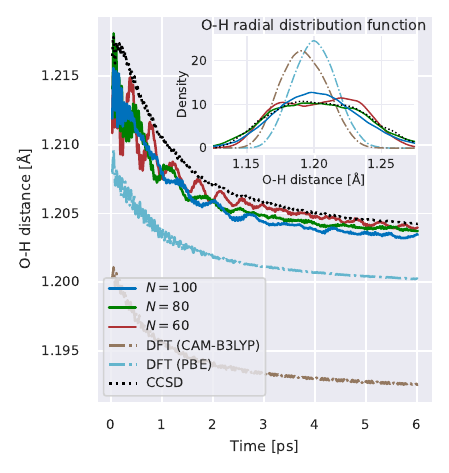}
        \caption{\vtwo{Thermalization to $298.15 \mathrm{K}$ of the trajectory-averaged distance between the central hydrogen and the oxygen atom of the Zundel cation. The main plot shows the mean distance as a function of propagation time obtained by interpolating from $N=60, 80, 100$ DMRG calculations, as well as DFT (with CAM-B3LYP~\cite{YANAI200451} and PBE exchange-correlation functionals) and CCSD trajectories. The mean corresponds to a running average of the distance between the atoms with a window of $100$ timesteps ($\approx 60.4 \, \mathrm{fs}$), and averaging over $500$ independent trajectories and both oxygen atoms. Each emsemble of trajectories required $5$ million energy and force calculations. Inset shows the thermalized radial distribution function of the oxygen from the central hydrogen, using a Gaussian smearing of individual data points in a kernel density analysis~\cite{chen2017tutorial,scikit-learn}, with a bandwidth of $\sigma=0.0025\,$\r{A}.}}
        \label{fig:Zundel_O_H_dist}
    \end{figure}

    \vtwo{Figure~\ref{fig:Zundel_O_H_dist} shows this thermalization in the average distance between the central hydrogen atom and the two oxygen atoms in the explored Zundel configurations. We find this statistically equilibrated distance to be converging to a slightly shorter length than CCSD as the number of training configurations is increased. An accurate description of this multi-center bond is key for the Grotthuss mechanism of proton transfer. The differences in these quantities are in stark contrast to the much shorter distances predicted by DFT MD simulations with two widely used exchange--correlation functionals, which indicate a more localized central hydrogen. We can observe this in the radial distribution function of the equilibrated configurations (inset) where the distribution is far flatter than the DFT methods, indicating an increased delocalization of the hydrogen between the water subunits. This is further corroborated by considering the magnitude of the dipole moment from the thermalized ensemble (see extended data), which we find decreases as the level of theory is increased from DFT to CCSD to the DMRG-interpolated configurations, indicating a preference for more symmetric distributions where the central hydrogen is delocalized and less bound at any instant to an individual oxygen atom.}

    \vtwo{The verifiably high-accuracy interpolation coupled with the high-accuracy DMRG training allows for validation in the use of CCSD for this system, with thermalized expectation values qualitatively in agreement. This is due to the lack of strong correlation in the explored molecular configurations. However, a significant advantage of this framework is the ability to also reliably converge the PES over the full phase space, including strongly correlated atomic configurations further from equilibrium where CCSD is unreliable and will potentially fail, including bond-breaking and transition state geometries.}
    
\begin{figure}
        \includegraphics{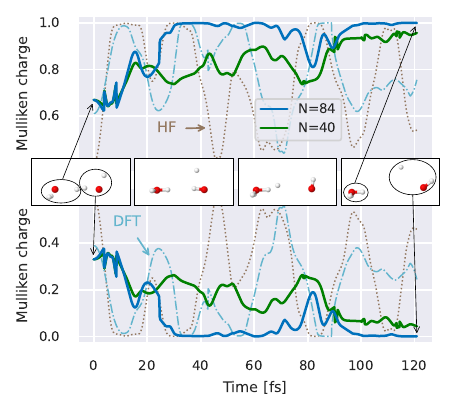}
        \caption{Predicted Mulliken charges for the hydronium (top panel) and water (bottom panel) sub-units from an MD simulation of the reaction from an eigenvector continuation with $N=40$ (green\vone{, dashed}) and $N=84$ (blue\vone{, solid}) DMRG data points. The system uses a 6-31G basis, with snapshots depicting the evolution of the molecular geometry at four evenly spaced times. Reference charges from simulations with DFT with a B3LYP exchange correlation (\vtwo{ligh blue, }dash-dotted) and HF (\vtwo{brown, }dotted) are included for comparison.}
        \label{fig:water_hydronium}
    \end{figure}

    \vtwo{To consider this scenario, we also} propagate a single high-energy trajectory within an NVE ensemble far from the Grotthuss mechanism dynamics, where the additional proton is inserted between the ions, \vtwo{interrupting the traditional hydrogen bond framework with a four-atom bridging bond as shown in the initial snapshot of} Fig.~\ref{fig:water_hydronium}. Increasing the number of DMRG training points to $N=84$, we are able to observe convergence in the specific short-time MD trajectory over the $120 \, \mathrm{fs}$ of the simulation (see extended data for evidence of this convergence with training data over the trajectory). Due to the fact that an explicit representation of the electronic state is retained over the trajectory, we also extract non-energetic electronic properties of the system over time. We use this to consider the evolution of the Mulliken charge as the electron density is redistributed around the system in response to the atomic motion, beyond the physics considered in traditional polarizable force fields.

    The positive charge is initially fairly evenly distributed amongst the water monomers, with the anticipation being that the water would rotate to adopt a lower-energy configuration. However, on the DMRG interpolated PES we find that before this is able to occur, a (neutral) hydrogen is ejected from the system leaving a bound hydronium and hydroxide radical system. The charges on the sub-units of this reaction show the redistribution of charge as the hydrogen oscillates a number of times before its eventual ejection from the system. This behavior is not seen on the more approximate HF or DFT electronic surfaces where the additional hydrogen remains bound over timescales enabling the water to rotate its orientation, with the HF charge distribution substantially in error even in the initial state. \vtwo{Comparing to CCSD across the $N=84$ trajectory, we find $\approx 40$ atomic configurations visited result in the CCSD energy diverging due to the presence of strong correlation effects, underlining the unreliability of the method for MD in these more unusual atomic conformations and transition states where strongly correlated electronic structure is found. This behavior is discussed further in the extended data, but underlines the applicability of the continuation across the phase space of the MD and the potential to describe dissociative dynamics~\cite{D4FD00062E}.}

    \vtwo{While these results demonstrate an effective acceleration scheme to converge the energy surface of this system to that of high accuracy methods, more consideration of the effects of basis size, nuclear quantum and solvent effects may be needed before predictions as to the nature of this physical reaction can be given with confidence \cite{10.1063/1.4941091,C4CP05192K}.} However, the fact that qualitative changes in dynamical behavior already result from the quality of the treatment of electronic correlation effects in the determination of the potential energy surface underlines the importance of a robust and systematically improvable approach to this electronic structure. The eigenvector continuation acceleration allows computation of this surface with high-level quantum chemical methods, and extends their scope to enable them to access timescales of atomic dynamics with provable convergence.

    \section*{Conclusion}\label{sec:conclusion}

    We develop a practical approach for eigenvector continuation of many-body electronic wave functions in {\em ab initio} settings. In contrast to the traditional paradigm of machine-learning force fields from training energies, this considers the interpolation of accurate wave functions across \vone{chemical space} \vtwo{the space of structural changes}, from which all electronic properties as well as atomic forces can be efficiently computed via a variational ansatz, avoiding the exponential complexity of the many-body states themselves. Using the scheme to converge the potential surface for molecular dynamics, we find examples of qualitatively different behavior to state-of-the-art techniques, demonstrating the importance of systematically converging the electronic structure across the timescales.

    The acceleration scheme therefore holds huge potential to extend the scope of modern highly accurate electronic structure to molecular dynamics applications. However, the potential of reliable wave function interpolation also goes beyond this, towards a consideration of non-adiabatic and beyond-Born--Oppenheimer effects, efficient geometry optimization for ground, transition states or conical intersections, as well as a general procedure for vibrations and phonons, raising the possibility of the routine extraction of thermodynamic variables from accurate quantum chemistry. The move from single-point electronic internal energies to (thermo)dynamical quantities within correlated electronic structure theory is a long saught-after ambition~\cite{https://doi.org/10.1002/anie.201709943}. The use of wave function interpolation with developments in solvers for the training data to extend system sizes could bring this closer to reality.

    \section*{Methods}
    \subsection*{\vtwo{Interpolating across the \emph{ab initio} potential energy surface}}
    At the core of the methodology lies the prediction of the ground state electronic energy for given molecular arrangement of $N_\mathrm{elec}$ electrons based on few exemplary solutions of the electronic structure problem at different molecular geometries.
    We define the {\em ab initio} electronic Hamiltonian for a $3 \times N_{\mathrm{nuc}}$ atomic configuration, $\mathbf{R}$, in a discrete basis of electronic orbitals $\{\chi(\mathbf{r};\mathbf{R})\}$ as~\cite{https://doi.org/10.1002/wcms.1123}
    \begin{align}
        \hat{H}(\mathbf{R}) & = \sum_{ij} h^{(1)}_{ij}(\mathbf{R}) \, \hat{c}^\dagger_{i} \hat{c}_{j} + \frac{1}{2} \sum_{ijkl} h^{(2)}_{ijkl}(\mathbf{R}) \, \hat{c}^\dagger_{i} \hat{c}^\dagger_{j} \hat{c}_{l} \hat{c}_{k} \nonumber \\
                            & \qquad + E_{\mathrm{nuc}}({\mathbf{R}}) \label{eq:hamiltonian}                                                                                                                                            \\
                            & = \sum_{ijkl} K_{ijkl}(\mathbf{R}) \, \hat{c}_i^\dagger \hat{c}_j^\dagger \hat{c}_l \hat{c}_k + E_\mathrm{nuc}({\mathbf{R}}) ,
    \end{align}
    with Fermionic creation and annihilation operators, $\hat{c}^\dagger$ and $\hat{c}$ acting on the orbitals\vtwo{, and $E_\mathrm{nuc}(\mathbf{R})$ the classical nuclear-nuclear repulsion energy}. 
    The one-electron terms, $h^{(1)}_{ij}(\mathbf{R})$, are matrix elements of the electron-nuclear and electronic kinetic operators, while the electron-electron repulsion integrals are
    \begin{align}
        h^{(2)}_{ijkl}(\mathbf{R}) & = \int \int d\mathbf{r}_1 d\mathbf{r}_2 \, \chi_i^*(\mathbf{r}_1; \mathbf{R}) \, \chi_j^*(\mathbf{r}_2;\mathbf{R}) \\ \nonumber
                                   & \qquad \frac{1}{|\mathbf{r}_1-\mathbf{r}_2|} \chi_k(\mathbf{r}_1;\mathbf{R}) \, \chi_l(\mathbf{r}_2;\mathbf{R}) ,  \\
                                   & = \langle ij|kl \rangle\vtwo{(\mathbf{R})} .
    \end{align}
    A convenient reduced two-body Hamiltonian which subsumes the one-body into the two-body term can be written as~\cite{PhysRevA.57.4219}
    \begin{align}
        K_{ijkl}\vtwo{(\mathbf{R})} &= \frac{1}{2} \langle ij|kl \rangle\vtwo{(\mathbf{R})} \\ \nonumber 
        &+ \frac{1}{2(N_\mathrm{elec} -1)}(\delta_{jl}h^{(1)}_{ik}\vtwo{(\mathbf{R})}+\delta_{ik}h^{(1)}_{jl}\vtwo{(\mathbf{R})}).
    \end{align}
    \vone{with $E_\mathrm{nuc}(\mathbf{R})$ the classical nuclear-nuclear repulsion term~\cite{PhysRevA.57.4219}}.

    As described in the main text, the eigenvector continuation proceeds via the definition of a symmetrically (Löwdin) orthonormalized atomic orbital basis (SAO)~\cite{10.1063/1.1747632, https://doi.org/10.1002/qua.981}. This allows the training wave functions to be transferred between the Hilbert spaces of different geometries by fixing their many-body probability amplitudes in this representation. These SAOs are defined with an orbital transformation of an underlying non-orthogonal atom-centered `AO' orbital basis set at each geometry, $\{\phi_\alpha(\mathbf{r};\mathbf{R})\}$, as
    \begin{equation}
        \chi_i(\mathbf{r};\mathbf{R}) = \sum_\alpha [\mathbf{S}(\mathbf{R})]^{-1/2}_{\alpha i} \, \phi_\alpha(\mathbf{r};\mathbf{R}).
        \label{eq:orbital_def}
    \end{equation}
    where $\mathbf{S}(\mathbf{R})$ is the atomic orbital overlap matrix
    \begin{equation}
        S_{\alpha \beta}(\mathbf{R}) = \int d \mathbf{r} \, \phi^\ast_\alpha(\mathbf{r};\mathbf{R}) \, \phi_\beta(\mathbf{r};\mathbf{R}).
    \end{equation}

    The continuation then proceeds according to the scheme outlined in the main text, with the evaluation of the transition two-body density matrices (t-2RDMs) and overlaps between the training points in their SAO representations. Of particular importance for molecular dynamics is the evaluation of analytic forces at each test geometry, which \vone{can be achieved via application of the Hellmann-Feynman theorem, which} \vtwo{is simplified due to the lack of response contributions from the many-body basis and} \vone{simple due to} the fully optimized variational nature of the interpolated states in the geometry-independent basis~\cite{PhysRev.56.340,10.1063/1.4927594}. This therefore only required the derivatives of the electron integrals in the AO basis~\cite{https://doi.org/10.1002/jcc.23981,https://doi.org/10.1002/wcms.1340, 10.1063/5.0006074}, as well as the derivative of the transformation from the atomic orbitals to the SAOs with respect to nuclear positions \vtwo{(a `Pulay force'~\cite{doi:10.1080/00268976900100941})}, which we evaluate via first order perturbation theory~\cite{bamieh2022tutorial}. The specifics of this evaluation is given in the Supplementary information.

    \subsubsection*{Approximate training data: DMRG}

    Rather than relying on exact (FCI) training data, we also consider modern numerically efficient approximations to the correlated electronic structure to allow for access to larger systems, which are nevertheless systematically improvable to the exact solution to the electronic Schrödinger equation for training. These require not only the evaluation of accurate many-body wave functions at the training geometries, but also the evaluation of the t-2RDMs and overlaps between different training states.

    Firstly, we consider the compression of the training wave functions in the form of \emph{Matrix Product States} (MPS), optimized via the \emph{density matrix renormalization group} (DMRG) algorithm~\cite{PhysRevLett.69.2863}. For this, we used the spin-adapted implementation from the \emph{block2} library~\cite{Block2,10.1063/5.0050902}, working directly in the Fock space defined by the SAOs.
    We optimize the training states with a schedule for exponentially increasing bond dimension (a factor of $1.8$ per increase) and decreasing noise in the MPS, a standard practice for stable {\em ab initio} DMRG~\cite{10.1063/1.4905329}, terminating when the difference of the energy upon fully relaxing the state at an increased bond dimension is less than a specified threshold, $\epsilon$.
    For all presented results, we set $\epsilon = 10^{-3} \, E_{\mathrm{h}}$ and start the MPS with an initial bond dimension of $34$, giving training data confidently below the accepted `chemical accuracy'. For the reference data for the hydrogen chain evolution (Fig.~\eqref{fig:H_chain_dimerization}), we set the tolerance to $\epsilon = 10^{-5} \, E_\mathrm{h}$, and starting MPS bond dimension to $61$.

    \subsubsection*{Approximate training data: CAS}

    In addition to the continuation from MPS training states optimized with DMRG, we also present the use of \emph{complete active space} (CAS) solvers to access the results of Fig.~\ref{fig:H2O_vibration}.
    These give an approximate ground state of the full electronic structure problem according to
    \begin{equation}
        |\Psi_\mathrm{CAS}\rangle = |0\rangle^{N_\mathrm{vir}} \otimes |\Psi_\mathrm{AS} \rangle \otimes |1\rangle^{N_\mathrm{core}},
    \end{equation}
    where $|\Psi_\mathrm{AS} \rangle$ represents the fully variationally optimized state over all many-electron configurations within a chosen active subspace of orbitals and electrons, while $|1\rangle^{N_\mathrm{core}}$ represents fully occupied orbitals spanning the remaining space of states that are occupied in a mean-field (in this case Hartree--Fock) description of the system, and $|0\rangle^{N_\mathrm{vir}}$ explicitly indicate that the higher-energy virtual states are unoccupied. In this way, the electronic fluctuations of a low-energy subspace are considered fully, with the choice of active space in this work selected purely based on the mean-field orbital energies about the chemical potential of the system.



    While this state can be straightforwardly optimized within a `CASCI' scheme implemented in the \emph{PySCF} package~\cite{https://doi.org/10.1002/wcms.1340, 10.1063/5.0006074}, we also require the evaluation of the overlap and the t-2RDMs between training states in their SAO basis, while the state is defined (and optimized) in a mean-field canonical basis. Therefore, it is necessary to rotate these many-body states into their respective SAO bases before the t-2RDMs and overlaps computed. We show this for the t-2RDM as
    \begin{align}
        \Gamma^{ijkl}_{ab} & =  \langle \Psi^{(a)}_\mathrm{CAS} | \hat{U}^\dagger_{\mathbf{R}^{(a)}}  \hat{c}^\dagger_i \hat{c}^\dagger_j \hat{c}_k \hat{c}_l \hat{U}_{\mathbf{R}^{(b)}} | \Psi^{(b)}_\mathrm{CAS} \rangle,
    \end{align}
    where $| \Psi^{(a/b)}_\mathrm{CAS} \rangle$ denotes the CAS states at the different training points and $\hat{U}_{\mathbf{R}^{(a/b)}}$ is the unitary transformation from the state in the basis of its canonical orbitals to the SAO basis for the corresponding training point.
    This is evaluated efficiently as a double summation over the {\em active space} many-electron configurations (including their core) of each training state
    \begin{multline}
        \Gamma^{ijkl}_{ab} = \\ \sum_{\mathbf{n}, \mathbf{n}' \in AS} C^{(a)\ast}_\mathbf{n} C^{(b)}_{\mathbf{n}'} \, \langle \mathbf{n} | \hat{U}^\dagger_{\mathbf{R}^{(a)}}  \hat{c}^\dagger_i \hat{c}^\dagger_j \hat{c}_l \hat{c}_k \hat{U}_{\mathbf{R}^{(b)}} | \mathbf{n}' \rangle,
        \label{eq:CAS_continuation}
    \end{multline}
    where $C^{(a/b)}_{\mathbf{n}/\mathbf{n}'}$ are the CASCI probability amplitudes of the active spaces.
    This single-particle unitary transformation $ \hat{U}^\dagger_{\mathbf{R}}$ can be formed as
    \begin{equation}
        \hat{U}_{\mathbf{R},ix} = \sum_{\alpha, \beta} Z_{\alpha i} \, S_{\alpha \beta} \, \tilde{Z}_{\beta x},
    \end{equation}
    where $Z_{\alpha i}$ is the transformation matrix from AO to SAO and $\tilde{Z}_{\beta x}$ is the transformation matrix from AO to canonical Hatree--Fock orbitals, while $S_{\alpha \beta}$ is the AO overlap matrix. All of these quantities are dependent on the specific training geometry, $\mathbf{R}$. 

    The inner products $\langle \mathbf{n} | \hat{U}^\dagger_{\mathbf{R}^{(a)}}  \hat{c}^\dagger_i \hat{c}^\dagger_j \hat{c}_k \hat{c}_l \hat{U}_{\mathbf{R}^{(b)}} | \mathbf{n}' \rangle$ from Eq.~\eqref{eq:CAS_continuation} can be identified as a matrix element between two different non-orthogonal Slater determinants~\cite{doi:10.1021/acs.jctc.5b01148}.
    The efficient evaluation of such overlaps between different non-orthogonal Slater determinants is discussed in Refs.~\cite{10.1063/5.0045442,10.1063/5.0122094}.
    We utilize the \emph{libgnme} package, together with its python interface \emph{pygnme}, to evaluate the overlaps and t-2RDMs between CAS states for the continuation in the SAO basis.
    Due to the non-orthogonality of the different CAS spaces, the double contraction of Eq.~\eqref{eq:CAS_continuation} results in a cost scaling quadratically in the size of the active space, thus more expensive than the evaluation of expectation values of a single point CAS state, however this cost could be reduced in the future by rotating to an intermediate basis representing the co-domain of the occupied CAS orbitals in a pair of CASCI training states.

    We note here that the eigenvector continuation scheme based on approximate solvers (especially CASCI) can result in inferred energies at training points which are variationally lower than those of the corresponding training energy at the same geometry. This is because it can mix contributions to the test state from other training states, with the resulting linear combination of approximate states lower in energy, resulting in an improved estimate for the energy at a test geometry than its corresponding training energy.
    Fig.~\ref{fig:H2O_energy_accuracy} of the extended data section exemplifies this behaviour, indicating lower predicted energies from the continuation of CAS states as compared to the reference CAS energies along the trajectory of a vibrating water molecule. \vtwo{In addition, discontinuities in CASCI potentials due to state crossings in the active space are possible and plague active space methods in general for complex trajectories. Due to the linearity of the ansatz in the interpolated energy surface, these discontinuities are necessarily removed, with an example for the water vibration of this phenomena shown in the extended data.}

    \subsubsection*{Gaussian Approximation Potentials}

    We include comparison results obtained from the prediction of potential energies via \emph{Gaussian Approximation Potentials} (GAP)~\cite{PhysRevLett.104.136403} -- a well-established framework for the prediction of potential energy surfaces and force fields.
    The model is extracted by fitting a data set of training geometries, $\{\mathbf{R}^{(a)}\}_{a=1}^N$, with associated energies $\{E^{(a)}\}_{a=1}^N$ using a kernel model~\cite{rasmussen2006gaussian} incorporating symmetries of atomic environments via the \emph{smooth overlap of atomic position} (SOAP) descriptors~\cite{doi:10.1021/acs.chemrev.1c00021}. We apply the GAP framework following standard approaches from the literature~\cite{C6CP00415F, PhysRevB.95.214302, 10.1063/5.0160898}, based on the implementation of the SOAP descriptors in the \emph{dscribe} package~\cite{HIMANEN2020106949}. Additional details of this prediction procedure can be found in the supplementary information.

    \subsection*{Eigenvector continuation for BOMD}

    The \vtwo{single-trajectory} Born--Oppenheimer molecular dynamics \vtwo{of Figs.~\ref{fig:H2O_vibration}, \ref{fig:H_chain_dimerization} and \ref{fig:water_hydronium}} were computed in vacuum based on a microcanonical (NVE) ensemble using the Velocity-Verlet integration implemented in \emph{PySCF}~\cite{https://doi.org/10.1002/wcms.1340, 10.1063/5.0006074,10.1063/1.442716,marx2000ab},
    according to the analytic nuclear gradients derived for the eigenvector continuation in the supplementary information.
    The nuclei in these simulations were initialized at rest, and chose a fixed timestep of $\delta t = 5 \, \mathrm{a.u.} \approx 0.121 \, \mathrm{fs}$ for the integration.

    \vtwo{To extract a thermalized ensemble for the dynamics of the Zundel cation of Fig.~\ref{fig:Zundel_O_H_dist}, we included a room temperate (298.15K) Berendsen thermostat~\cite{berendsen1984molecular} as implemented in \emph{PySCF} to obtain a canonical (NVT) ensemble of trajectories. This scheme relies on an additional rescaling of the velocities after each integration step to achieve an exponential convergence to the target temperature with a timescale $\tau$. Initial velocities for each trajectory were drawn from a Maxwell-Boltzmann distribution, while the nuclei positions were initialized in the ground state geometry obtained from CCSD(T) in a large basis set from Ref.~\cite{10.1063/1.1834500}. The dynamics proceeded with a total of $10,000$ integration steps with $\delta t = 25 \, \mathrm{a.u} \approx 0.605 \, \mathrm{fs}$, and a thermalization time constant of $\tau = 250 \, \mathrm{a.u.} \approx 6.05 \, \mathrm{fs}$. This required $5 \times 10^6$ force calculations to propagate the ensemble of $500$ trajectories over the $6 \, \textrm{ps}$ timescale considered.}

    \subsubsection*{Active learning for data selection}

    For the molecular dynamics applications, we perform an active learning scheme in which we identify and select suitable molecular configurations for training the eigenvector continuation scheme on-the-fly.
    This scheme is based on iteratively running the MD with a given training set, and selecting an enlarged training dataset with a new molecular configuration from the sampled trajectory. A correlated electronic structure calculation is performed at the selected geometry which is then included in the training data for an improved inferred potential energy surface and resulting MD trajectory in the next step.
    Starting from just a single training state (the initial geometry) and iteratively adding new configurations to the dataset in this way, the number of costly electronic structure calculations can be minimized and the trajectory can be systematically and rapidly converged, noting that adding training geometries from the simulated trajectories guarantees an improved prediction in each step.


    To select the new training geometry, we develop a `distance' heuristic for all geometries along the trajectory, which can be used as a metric for the addition of the data, and quantifies the suitability of the current training data in describing the test state at each point. The point along the trajectory with the largest measure is added to the training data set. Since the (non-degenerate) ground states along the trajectory are uniquely defined by the {\em ab initio} Hamiltonian at each geometry, we use the differences between the Hamiltonian elements at the training points and all trajectory points in defining this measure. Defining these elements in their respective SAO basis of each geometry used for the inference also ensures that the invariances and symmetries of the eigenvector continuation are also respected in this measure. Specifically, we define this Hamiltonian distance between two geometries, $D(\mathbf{R}, \mathbf{R}')$, as
    \begin{multline}
        D(\mathbf{R}, \mathbf{R}') = \sum_{ij} |h^{(1)}_{ij}(\mathbf{R}) - h^{(1)}_{ij}(\mathbf{R}')|^2 + \\ \frac{1}{2} \sum_{ijkl} |h^{(2)}_{ijkl}(\mathbf{R}) - h^{(2)}_{ijkl}(\mathbf{R}')|^2. \label{eq:ham_dist}
    \end{multline}

    In addition to respecting the symmetries of the model, this ensures that two geometries with similar Hamiltonians (and thus wave functions) are considered similar, even though an evaluation of the Euclidean distance between these two geometries might be large (e.g., for geometries from near a dissociated limit).
    To add a new configuration, we evaluate $D(\mathbf{R}(t), \mathbf{R}^{(a)})$ for all geometries $\mathbf{R}(t)$ from the trajectory and each training geometry $\mathbf{R}^{(a)}$ already contained in the training set.
    We then pick that configuration $\mathbf{R}(t_{add})$ from the trajectory for which the distance to the closest training configuration is maximal, i.e., where
    \begin{equation}
        \quad t_\mathrm{add} = \argmax_t \left(\minimum_a (D(\mathbf{R}(t), \mathbf{R}^{(a)})) \right)
    \end{equation}

    \vtwo{To gauge the systematic convergence of the NVE MD single-shot trajectories, we can track the variational lowering (and hence improvement) of the potential energy surface as the dataset is enlarged. This is done by comparing the PES from the two data set sizes along the same trajectory corresponding to the larger of the two data sets.
    Exploiting the variationality of the method}, it is guaranteed that the potential energy inference with the larger dataset will be lower or equal to the predictions with the smaller dataset, and we use the difference between the predicted energies as a convergence measure.
    In our applications, we terminate the simulation when the predicted energy with the enlarged data set stays within a tolerance of $\epsilon = 10^{-3} \, E_{\mathrm{h}}$ along the full MD trajectory for two iterations in a row. Examples of this convergence are shown in the extended data.

    \vtwo{To manage the increased data volume when generating the statistical canonical ensemble of trajectories for the NVT Zundel cation results of Fig.~\ref{fig:Zundel_O_H_dist}, we use a somewhat coarser scheme to select the training configurations. We start with just the initial configuration in the training set, and randomly sub-sample $100$ trajectories from the ensemble of $500$ trajectories generated by the prior CAM-B3LYP DFT dynamics. For all timesteps of these $100$ trajectories, the Hamiltonian distance metric of Eq.~\ref{eq:ham_dist} is computed and the $19$ geometries with the largest value of this metric are identified for inclusion in an enlarged training data set. The continuation scheme is then run for $500$ NVT trajectories with these $20$ training points. We compute the Hamiltonian metric along the full path of a new random selection of $100$ of these inferred trajectories in order to identify a further set of $20$ geometries to perform explicit DMRG calculations to iteratively enlarge the training data set until the desired size is reached. This training set is taken to be the same for all trajectories in an ensemble. It should be stressed that only the first batch of $19$ geometries are taken from the DFT-derived trajectories, after which subsequent batches of training geometries are found self-consistently to ensure a systematically reducing bias due from the DFT paths.}

    \section*{Code and data availability}

    The code and inputs to fully reproduce the numerical experiments of this work can be found at \url{https://github.com/BoothGroup/evcont}.
    Furthermore, we provide the videos of the \vtwo{isolated NVE} molecular dynamics trajectories presented in this work at \url{https://doi.org/10.5281/zenodo.10658975}.

    \section*{Acknowledgments}

    We thank Huanchen Zhai for support with the \emph{block2} code, Hugh Burton for support with the \emph{libgnme} code, Carlos Mejuto-Zaera for helpful insights about an independently-developed related scheme, Oliver Backhouse for technical help and Kemal Atalar, Venkat Kapil and Lachlan Lindoy for additional feedback on the manuscript.
    We gratefully acknowledge support from the Air Force Office of Scientific Research under award number FA8655-22-1-7011 and the UK Materials and Molecular Modelling Hub for computational resources, which is partially funded by EPSRC (EP/T022213/1, EP/W032260/1 and EP/P020194/1). YR also acknowledges the support of the Engineering and Physical Sciences Research Council [grant EP/Y005090/1].

    \section*{Author contributions}

    YR and GHB jointly developed the methodology and wrote the manuscript.
    YR implemented the approach and performed the numerical experiments. GHB supervised the project.


    \section*{Extended data}

    \subsubsection*{Accuracy of the potential energy surface for water vibration and hydrogen chains}

\begin{figure}
        \includegraphics{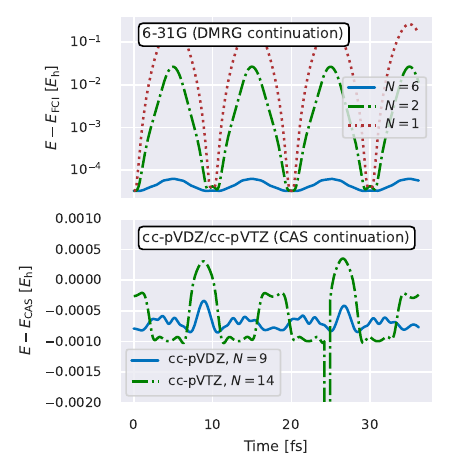}
        \caption{Difference between the predicted energy and a reference method over all geometries from the MD trajectory of a water molecule. Top panel: Error compared to FCI of DMRG-based eigenvector continuation with increasing data set in the 6-31G basis. Bottom panel: Difference between CASCI(4,8)-trained eigenvector continuation and CASCI(4,8) energies at each geometry along the trajectory for $N=9$ training points in a cc-pVDZ basis, and $N=14$ training points in a cc-pVTZ basis. We note that variationality with respect to this approximate training data is not expected.}
        \label{fig:H2O_energy_accuracy}
    \end{figure}

    In Fig.~\ref{fig:H2O_energy_accuracy}, we report the difference between the inferred and reference energies at every point along the simulated molecular dynamics trajectory of the water vibration model in the main text.
    The top panel shows the accuracy of the potential energy surface with the 6-31G basis set, where we performed the DMRG-based continuation scheme, compared to the exact potential energy surface.
    It can be seen that increasing the number of training states rapidly converges the full trajectory with $N=6$ training states achieving an accuracy well below $10^{-4} \, E_{\mathrm{h}}$ across all points.

\begin{figure}
        \includegraphics{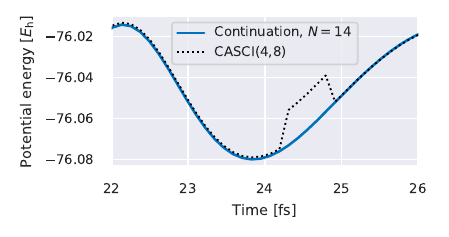}
        \caption{\vtwo{Potential energy surface along a short time interval from the MD trajectory of the water molecule in a cc-pVTZ basis from Fig.~\ref{fig:H2O_energy_accuracy}. The reference CASCI(4,8) potential energy is shown along with the $N=14$ interpolated surface based on selected data points of this reference. Significantly, the chosen part of the surface exhibits a discontinuity due to the inadequacy of the active space choice which is alleviated by the interpolation scheme.}}
        \label{fig:H2O_vtz_PES}
    \end{figure}

    The variationality of the method guarantees that the predicted energies are always an upper bound to the {\em exact} ground state energy at any geometry.
    Nonetheless, when the continuation is based on {\em approximate} training wave functions, the computed linear combination of training states may result in an energy lower than the reference training energy at that geometry.
    This is exemplified in the bottom panel of Fig.~\ref{fig:H2O_energy_accuracy}, detailing the energetic difference between the prediction and CASCI energies used for the training data along the trajectory in cc-pVDZ and cc-pVTZ basis sets.
    While the same active space sizes were used for the training states as for the computation of the reference energies, the inferred energies generally lie below the CASCI reference energies.
    Although this improvement is small, mostly less than $1 \, mE_{\mathrm{h}}$, it is obtained for the majority geometries from the converged trajectory, ultimately improving the accuracy beyond what is obtained with the reference method.
    \vtwo{As a further noteworthy difference between the CASCI and CASCI-trained model, for a small path of the trajectory corresponding to times between $24$ and $25 \, \mathrm{fs}$ evolution time in the cc-pVTZ basis of Fig.~\ref{fig:H2O_energy_accuracy}, a much more significant improvement of the continuation results over the reference method becomes apparent.
    As is shown in Fig.~\ref{fig:H2O_vtz_PES}, this is due to a discontinuity in the CASCI ground state potential surface caused by a change of states included in the active space for these geometries in the trajectory. A more careful choice of active space is likely to have alleviated this problem, but we highlight it here since it is clear that this discontinuous change does not affect the interpolated surface, which is necessarily smoothly changing with geometry, and therefore also mitigates a significant challenge of the use of active space methods in molecular dynamics.}

\begin{figure}
        \includegraphics{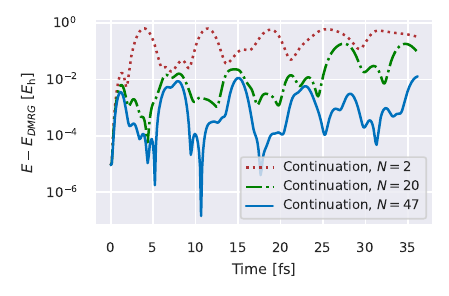}
        \caption{Energy difference between reference DMRG energies and the DMRG-trained eigenvector continuation for the trajectory of the $30$-atom hydrogen chain (main results presented in Fig.~\ref{fig:H_chain_dimerization}), with $N=2$, $N=20$, and $N=47$ training data points.}
        \label{fig:H30_energy_accuracy}
    \end{figure}

    Analogously to Fig.~\ref{fig:H2O_energy_accuracy}, Fig.~\ref{fig:H30_energy_accuracy} depicts the difference between the energy predicted by the eigenvector continuation and the reference DMRG energy for the converged MD trajectory of the $30$-atom hydrogen system of the main text.
    It can be seen that, even though there is no prior knowledge about the targeted trajectory, the on-the-fly dataset generation for the molecular dynamics simulation achieves a potential energy surface resolved to roughly a few millihartree along the whole trajectory. \vtwo{The fluctuations in this error, often to close to zero, reflect the distance of the configuration explored in the trajectory to the training geometries. Since the training data is essentially exact, the inferred model must also be necessarily exact at these points, leading to the decreases to zero in the log of the error near these parts of the phase space in the trajectory.}

    \subsubsection*{Validation of training for thermalized Zundel cation ensembles}

        \vtwo{While it is possible to exploit the variationality of the approach to demonstrate convergence via systematic lowering of the energy across the trajectory, for the dynamics of the NVT ensemble of nuclear trajectories for the Zundel system of Fig.~\ref{fig:Zundel_O_H_dist}, we validate the accuracy of the inference via comparison to additional explicit DMRG calculations along the generated trajectories that are not in the training set.
        We select the geometries for this validation data set by sampling a total of $1000$ configurations from all the trajectories (generated from $N=100$ training data points) at equidistant time intervals along the evolution.}

\begin{figure}
        \includegraphics{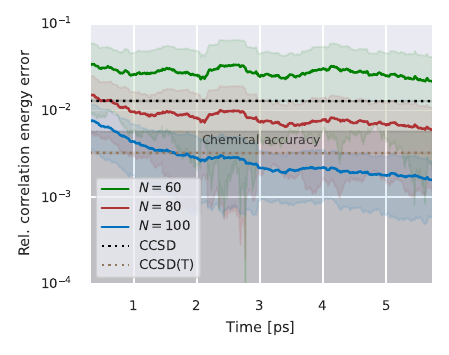}
        \caption{\vtwo{Relative correlation energy error for the ensemble of Zundel cation trajectories with different numbers of training geometries ($N$), taken from a sub-sampling of $1000$ representative geometries from equidistant times along the final $N=100$ trajectory. This quantifies the test error across the trajectories against additional reference DMRG calculations not used in their training. The plot depicts the running average over $10$ out of the total of $1000$ test geometries, with the corresponding standard deviation highlighted by the shaded areas, and `chemical accuracy' denoted by the gray shaded area -- an accuracy for which the $N=100$ model clearly surpasses for the thermalized dynamics.}}
        \label{fig:energy_validation_Zundel}
    \end{figure}

    \vtwo{Figure~\ref{fig:energy_validation_Zundel} shows the relative correlation energy error from the prediction of the continuation model with $N=60, 80$ and $100$ training data sizes, compared to $1000$ exact reference points obtained via DMRG for this validation set over the trajectory.
        The solid line represents a running average over ten test geometries, with additional shading indicating the standard deviation. The consistent reduction in the energy error as more data points are used for the prediction is evident.
        For a training set with $N=100$ configurations, we obtain a mean correlation energy error below chemical accuracy for the thermalized system, also demonstrating an improved accuracy compared to both CCSD and CCSD(T), increasing our confidence in the fidelity of the predicted thermodynamic quantities.}

    \subsubsection*{Thermalized Zundel cation dipole moment}
\begin{figure}
        \includegraphics{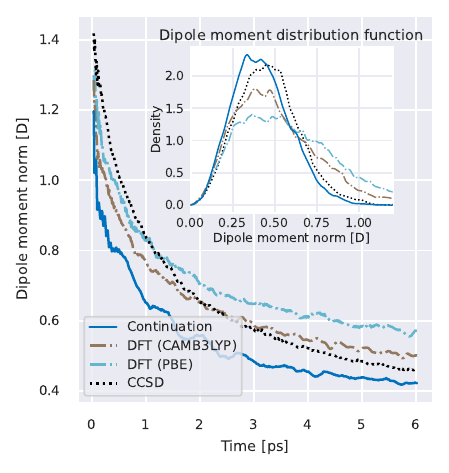}
        \caption{\vtwo{Dipole moment norm as a function of propagation time averaged over the ensemble of $500$ trajectories for the same setup as discussed in Fig.~\ref{fig:Zundel_O_H_dist}. As in Fig.~\ref{fig:Zundel_O_H_dist}, the main panel depicts a running average over $100$ propagation timesteps, and the inset shows the thermal distribution function obtained via a kernel density analysis with Gaussian smearing with $\sigma=0.01 \, \mathrm{D}$ over the geometries of the last $100$ iterations. Continuation results correspond to the $N=100$ interpolation model.}}
        \label{fig:dipole_norm_Zundel}
    \end{figure}

    \vtwo{In addition to the statistical convergence of the oxygen-to-hydrogen distance for the thermalized ensemble shown in Fig.~\ref{fig:Zundel_O_H_dist} of the main text, we additionally show the convergence of an electronic quantity of interest.
        The main panel of Fig.~\ref{fig:dipole_norm_Zundel} shows the averaged norm of the dipole moment over the thermalized ensemble of trajectories (with respect to the center of mass of the system) as a function of the propagation time, with the inset displaying the associated thermalized distribution function at the final time. As a response property, this is likely to be a more sensitive quantity with respect to the quality of the electronic structure over the trajectory. While the DMRG-trained interpolation (with $N=100$) gives a flatter thermalized distribution function for the Oxygen-Hydrogen distance compared to DFT (see Fig.~\ref{fig:Zundel_O_H_dist}), we find that the magnitude of the dipole moment is reduced compared to DFT (both with CAM-B3LYP and PBE exchange correlation functionals) and CCSD, both for the final thermalized ensemble, and for the full trajectory. Density functional theory in particular gives a flatter and more skewed thermalized distribution compared to both CCSD and the continuation.
    }

    \subsubsection*{Zundel cation energy surface convergence for high-energy trajectory}

\begin{figure}
        \includegraphics{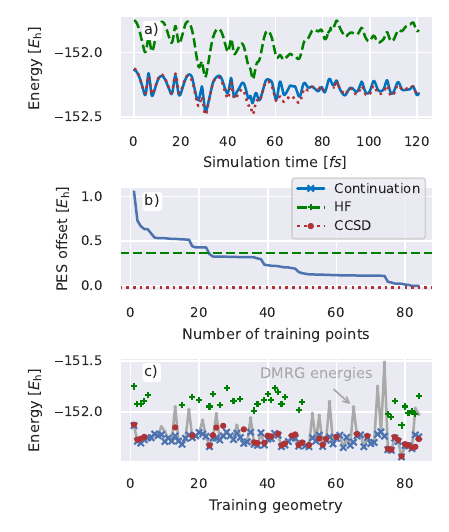}
        \caption{Characteristics of potential energy surfaces and single point training energy estimates for the simulation of the water-hydronium reaction from the eigenvector continuation (blue), Hartree-Fock (green), as well as CCSD (red). The top panel (a) shows the potential energy for these methods as a function of time over the $N=84$ converged MD trajectory. Panel (b) shows the average energy over the final trajectory which systematically converges as more training data points are included in the eigenvector continuation. Panel (c) shows the single point energy estimates for each of the $N=84$ training geometries. Blue crosses denote the energy of the DMRG-trained eigenvector continuation at the training geometries, the gray line gives the single point energy estimates from the individual DMRG calculations used for the training points (noting the spikes corresponding to poor DMRG convergence), while red dots and green crosses show the CCSD and HF values at the training geometries which were able to be converged along the trajectory with default parameters.}
        \label{fig:PES_water_hydronium}
    \end{figure}

    Figure~\ref{fig:PES_water_hydronium} demonstrates the convergence of the DMRG-trained potential energy surface (PES) underpinning the \vtwo{high-energy} Zundel cation \vtwo{NVE trajectory of Fig.~\ref{fig:water_hydronium}.}
    In the top panel, we show the final $N=84$ inferred electronic energy along the trajectory, compared to approximate Hartree--Fock (HF) and coupled cluster with single and double excitations (CCSD). The obtained values improve significantly upon single point energy estimates from Hartree-Fock, and are generally comparible with CCSD over the whole simulation, noting that CCSD is not a variational method and therefore can be above or below the exact energy. Nevertheless, we expect CCSD to be generally an accurate method to describe this system over many of the geometries visited. \vtwo{However, it is significant that for approximately $40$ geometries along the $N=84$ trajectory, CCSD could not converge with reasonable simulation and convergence parameters, indicating that more strongly correlated electronic structure is found along the trajectory. In contrast, the eigenvector continuation, which lacks any self-consistent or iterative aspects in inferring the electronic structure for each geometry (even at a mean-field level), can be run in an entirely robust and black-box fashion, with the ability to describe strongly correlated geometries only dependent on the solver used for the training.}

    In the center panel of the figure we highlight the convergence of the PES as a function of the number of training geometries used, showing the average electronic energy over the final trajectory for different numbers of training points (shifting the energies such that the $N=84$ point has a zero average PES). While the variational character of the continuation guarantees a monotonic decrease of this average potential energy as more training configurations are added, various flat regions can be observed, indicating training configurations not contributing to a substantial improvement to the potential energy of the final trajectory. This highlights the potential for further improvements to the data selection scheme in order to identify the most appropriate training configurations and further accelerate the convergence of the PES with training points.

    Finally, the bottom panel of Fig.~\ref{fig:PES_water_hydronium} shows the energy at the $84$ training configurations used to train the model in the final converged MD simulation.
    At a number of these training geometries, the HF and/or CCSD were not able to fully converge, and these points have been omitted. While optimizing convergence parameters may enable convergence, this is potentially a flag for stronger correlation effects at these points, where the DMRG-trained continuation would be more valuable and HF and CCSD unsuitable. Furthermore, there are clear indications of at least 13 training points where even the DMRG training data was not able to correctly converge, likely due to getting stuck in local minima in the DMRG sweep algorithm that optimizes the MPS.

    This also highlights the lack of `black-box' character for these highly accurate wave function based electronic structure methods, which means that much care is needed if they were to be applied to MD on their own. Likely the convergence of the continuation with respect to training points would have been improved had these been fully converged, however it is noteworthy that this does not substantially impact upon the quality of the inference of the eigenvector continuation at these training points, which is found to be substantially more accurate (variationally lower) than their corresponding training point at these points even where the optimization of the training wave function failed. \vtwo{This is analogous to the robustness of the interpolation demonstrated in Fig.~\ref{fig:H2O_vtz_PES} for CASCI training in the presence of discontinuities. Similarly}, the surface for the MD trajectory necessarily remains smooth and the gradients well-behaved. We note that the nature of the active data selection means that the molecular geometries of these training points are unlikely to actually feature in the final MD trajectory, since each time the data set is enlarged, the PES changes and the MD follows a different path. In the final MD simulation, the HF does converge for all points along the trajectory, and CCSD with default optimization parameters fails $\sim 4\%$ of the time.

    We note finally that the rationale for selection of additional data points via the Hamiltonian metric of Eq.~\ref{eq:ham_dist} builds upon the assumption that data points add the correct training state to the dataset and therefore significantly improve the potential energy surface at that geometry. Adding data points for which an appropriately converged solution to the ground state was not obtained (in this case with DMRG) can thus negatively affect the performance of the data selection for subsequent runs and the rate of convergence of the method to exactness. Improving the convergence of the molecular dynamics simulation even in limits where the data set comprises inaccurate solutions to the electronic structure problem remains a subject of future research.

\end{bibunit}

\ifsuppinfo
    \clearpage
    \newpage
\begin{bibunit}
    
    \section*{Supplementary information}


    \subsubsection*{Nuclear forces for eigenvector continuation}

    Based on the potential energy estimate $E(\mathbf{R})$ for a nuclear configuration $\mathbf{R}$, the force can be defined as
    \begin{equation}
        F(\mathbf{R}) = - \frac{\partial E(\mathbf{R})}{\partial \mathbf{R}} = - \left(\frac{\partial E_\mathrm{nuc}(\mathbf{R})}{\partial \mathbf{R}} + \frac{\partial E_\mathrm{el}(\mathbf{R})}{\partial \mathbf{R}}\right) ,
        \label{eq:force_field}
    \end{equation}
    which as shown can be split up into the electronic and nuclear contributions to this force.
    Focusing on the electronic contribution from the eigenvector continuation (and dropping the subscript for clarity), we apply the Hellmann-Feynman theorem~\cite{PhysRev.56.340}, within the restricted subspace spanned by the training states, giving
    \begin{equation}
        \frac{\partial E(\mathbf{R})}{\partial \mathbf{R}} = \frac{\partial}{\partial \mathbf{R}} \left( \boldsymbol{\psi}^\dagger \, \boldsymbol{\mathcal{H}} \, \boldsymbol{\psi} \right) =  \boldsymbol{\psi}^\dagger \, \frac{\partial \boldsymbol{\mathcal{H}}}{\partial \mathbf{R}} \, \boldsymbol{\psi},
    \end{equation}
    where $\boldsymbol{\psi}$ is the inferred ground-state eigenfunction from the diagonalization of the Hamiltonian in the many-body basis of the fixed training states (Eq.~2 of the main text).
    Substituting in the definition of the Hamiltonian from Eq.~5, similar to the derivation in Ref.~\cite{10.1063/1.4927594}, we obtain
    \begin{equation}
        \frac{\partial E(\mathbf{R})}{\partial \mathbf{R}} = \sum_{ijkl} \frac{\partial K_{ijkl}(\mathbf{R})}{\partial \mathbf{R}}  \, \Gamma^{ijkl} 
    \end{equation}
    where $\Gamma^{ijkl}$ is the two-electron reduced density matrix of the inferred state,
    \begin{equation}
        \Gamma^{ijkl} = \sum_{a,b}^N \psi^*_a \, \Gamma^{ijkl}_{ab} \, \psi_b.
    \end{equation}
    The reduced Hamiltonian gradient, $\frac{\partial K_{ijkl}(\mathbf{R})}{\partial \mathbf{R}}$, is obtained according to
\begin{multline}
    \frac{\partial K_{ijkl}(\mathbf{R})}{\partial \mathbf{R}} = \frac{1}{2} \frac{\partial h^{(2)}_{ijkl}(\mathbf{R})}{\partial \mathbf{R}} +\\ \frac{1}{2(N_\mathrm{elec} -1)} \left(\delta_{jl} \frac{\partial h^{(1)}_{ik}(\mathbf{R})}{\partial \mathbf{R}} +\delta_{ik}\frac{\partial h^{(1)}_{jl}(\mathbf{R})}{\partial \mathbf{R}}\right)
\end{multline}

    The evaluation of the the one- and two-electron integral gradients in the SAO basis requires the derivative of the transformation $\mathbf{Z}(\mathbf{R})$ from the atomic orbitals to the SAO basis (a Pulay-like force contribution), as well as the gradients of the integrals between atomic orbitals.
    For the one-electron integral derivative, we obtain
    \begin{multline}
        \frac{\partial h^{(1)}_{ij}}{\partial \mathbf{R}} = \sum_{\alpha \beta} Z_{\alpha i} \, Z_{\beta j} \, \frac{\partial \tilde{h}^{(1)}_{\alpha \beta}}{\partial \mathbf{R}} \\ + \sum_{\alpha \beta} \frac{\partial Z_{\alpha i}}{\partial \mathbf{R}} \, Z_{\beta j} \, \tilde{h}^{(1)}_{\alpha \beta} + \sum_{\alpha \beta} \frac{\partial Z_{\beta j}}{\partial \mathbf{R}} \, Z_{\alpha i} \, \tilde{h}^{(1)}_{\alpha \beta},
    \end{multline}
    where greek indices refer to atomic orbitals, latin indices to SAO, and $\tilde{h}^{(1)}_{\alpha \beta}$ denotes the one-electron integral in the atomic orbital basis.
    The derivatives of the integrals over the atomic orbitals, are readly accessible from standard libraries, and we utilize the implementation in the \emph{Libcint} library~\cite{https://doi.org/10.1002/jcc.23981} as bundled in the \emph{PySCF} package~\cite{https://doi.org/10.1002/wcms.1340, 10.1063/5.0006074}.

    The gradient of the SAO transformation, $ \frac{\partial Z_{\alpha i}}{\partial \mathbf{R}}$, follows directly from the definition of the Löwdin orthogonalization, and application of the chain rule.
    This gives
    \begin{equation}
        \frac{\partial Z_{\alpha i}}{\partial \mathbf{R}} = \sum_{\beta \gamma} \frac{\partial S_{\beta \gamma}}{\partial \mathbf{R}} \frac{\partial (\mathbf{S}^{-1/2})_{i \alpha}}{\partial S_{\beta \gamma}}.
    \end{equation}
    In addition to the derivative of the overlap matrix between atomic orbitals, this requires the evaluation of the derivative of the inverse square root matrix elements with respect to the overlap elements, $\frac{\partial (\mathbf{S}^{-1/2})_{i \alpha}}{\partial S_{\beta \gamma}}$.
    To evaluate this derivative, we consider the spectral decomposition of $\mathbf{S}$ according to
    \begin{equation}
        \mathbf{S} = \mathbf{T} \, \mathbf{s} \, \mathbf{T}^\dagger, \label{eq:ovlp_spectrum}
    \end{equation}
    from which we obtain
    \begin{multline}
        \frac{\partial \mathbf{S}^{-1/2}}{\partial S_{\beta \gamma}} = \frac{\partial \mathbf{T}}{\partial  S_{\beta \gamma}} \, \mathbf{s}^{-1/2} \, \mathbf{T}^\dagger + \\ \mathbf{T} \, \frac{\partial \mathbf{s}^{-1/2}}{\partial  S_{\beta \gamma}} \, \mathbf{T}^\dagger + \mathbf{T} \, \mathbf{s}^{-1/2} \, \frac{\partial \mathbf{T}^\dagger}{\partial  S_{\beta \gamma}}.
    \end{multline}

    The eigenvector and eigenvalue derivatives can be evaluated via first order perturbation theory of each element of $S_{\beta \gamma}$ and evaluating the derivative at vanishing perturbation strength~\cite{bamieh2022tutorial}.
    The chain rule then provides
    \begin{equation}
        \frac{\partial s^{-1/2}_i}{\partial  S_{\beta \gamma}} = - \frac{1}{2} \, s^{-3/2}_i \, \frac{\partial s_i}{\partial  S_{\beta \gamma}} = - \frac{1}{2} \, s^{-3/2}_i \, T^\ast_{\beta i} \, T_{\gamma i},
    \end{equation}
    and we obtain for the derivative of the eigenvector coefficients
    \begin{equation}
        \frac{\partial T_{\alpha i}}{\partial  S_{\beta \gamma}} =  - \sum_{j \neq i} \frac{T^\ast_{\beta j} \, T_{\gamma i} \, T_{\alpha j}}{s_j - s_i}.
        \label{eq:eigenvec_derivative}
    \end{equation}
    The sum in the above expression runs over all eigenstates except $i$.
    Additional care needs to be taken if eigenvalue degeneracies are present in Eq.~\eqref{eq:ovlp_spectrum}, in which case we appeal to standard degenerate perturbation theory~\cite{bamieh2022tutorial}, ensuring we rotate the eigenvectors in each degenerate subspace such that the applied perturbation in this degenerate subspace is diagonal.
    In this case, Eq.~\eqref{eq:eigenvec_derivative} becomes valid by only summing over all eigenstates which have a distinct eigenvalue from $s_i$.

    \subsubsection*{Details of GAP model predictions}
    \label{sec:GAP_SI}
    Within the application of the GAP framework, the potential energy for a test geometry is predicted based on a kernel model according to
    \begin{equation}
        E(\mathbf{R}) = \sum_{a=1}^N w_a \, k(\mathbf{R}, \mathbf{R}^{(a)}),
        \label{eq:GAP_prediction}
    \end{equation}
    where the weights $w_a$ may be interpreted as learned parameters of the model, and the kernel function $k$ defines a similarity measure between nuclear geometries.

    Without fully optimizing the various design choices, we follow standard approaches to construct a suitable kernel function within the GAP framework based on the idea of a smooth overlap of atomic positions (SOAP)~\cite{C6CP00415F}.
    This representation defines a set of features for each atom in the system based on its local atomic environment represented in a basis of spherical harmonics together with radial basis functions automatically incorporating important symmetries into the representation.
    We utilize the implementation of the SOAP features from the \emph{dscribe} library~\cite{HIMANEN2020106949}, implementing the SOAP features as described in Ref.~\cite{C6CP00415F}.
    Without additional hyperparameter tuning, we constructed the SOAP features in our tests by going up to 20\textsuperscript{th} order in the spherical harmonics and 10\textsuperscript{th} order in the radial basis functions, and chose a radial cutoff of $10 \, \text{\AA}$.

    Based on these SOAP features, we construct an averaged kernel function~\cite{C6CP00415F}, given as the average of scalar products between the SOAP feature vectors over all pairs of atoms in the system.
    Denoting the vector of SOAP features for the $i$-indexed atom as $\mathbf{p}(\mathbf{R}_i)$, this kernel, appropriately normalized, is thus given as
    \begin{equation}
        k(\mathbf{R}, \mathbf{\mathbf{R}}') = \frac{\sum_{i j} \mathbf{p}(\mathbf{R}_i) \cdot \mathbf{p}(\mathbf{R}'_j)}{\sqrt{\mathcal{N}(\mathbf{R}) \times \mathcal{N}(\mathbf{R}')}},
        \label{eq:kernel}
    \end{equation}
    with a normalization constant
    \begin{equation}
        \mathcal{N}(\mathbf{R}) = \sum_{i j} \mathbf{p}(\mathbf{R}_i) \cdot \mathbf{p}(\mathbf{R}_j).
    \end{equation}

    Following standard Gaussian process regression techniques~\cite{rasmussen2006gaussian}, the weights of the model are obtained by inversion of the kernel matrix according to
    \begin{equation}
        \mathbf{w} = (\mathbf{K} + \sigma^2 \mathbb{1})^{-1} \mathbf{E}.
    \end{equation}
    Here, $\mathbf{w}$ denotes the vector of weights for all data points, $\mathbf{K}$ is the matrix of pairwise kernel values between the training configurations, and $\mathbf{E}$ is a vector comprising the training energies.
    For the results shown in Fig.~2 of the main text, we set the additional noise parameter $\sigma^2$, effectively regularizing the fit, to a small fixed value of $\sigma^2=10^{-15}$, without additional optimization.

    While it is possible to train GAP models also on force data to improve the prediction~\cite{PhysRevB.95.214302}, here we simply estimate the force by differentiating through the potential energy predictor according to Eq.~\eqref{eq:GAP_prediction}.
    The force prediction is therefore obtained as
    \begin{equation}
        F(\mathbf{R}) = - \sum_{a=1}^N w_a \frac{\partial k(\mathbf{R}, \mathbf{R}^{(a)})}{\partial \mathbf{R}},
    \end{equation}
    with the kernel function as defined in Eq.~\eqref{eq:kernel}.
    It should be noted that the SOAP features used for the kernel are atom-centered, and therefore the derivative with respect to these basis functions needs to be taken into account when evaluating the gradient, as implemented in the \emph{dscribe} library~\cite{HIMANEN2020106949}.



\end{bibunit}


\fi

\end{document}